\documentstyle[eqsecnum,aps,epsf]{revtex}

\def\ep{\varepsilon}

\def\cm#1{}

\begin{document}
\setcounter{figure}{0}
\Roman{figure}

\title{{
Critical Exponents
from
Five-Loop
Strong-Coupling
$\phi^4$-Theory
in $4- \ep$ Dimensions.
}}
\author{Hagen Kleinert%
 \thanks{Email: kleinert@physik.fu-berlin.de 
URL:
http://www.physik.fu-berlin.de/\~{}kleinert \hfil
}}
\address{Institut f\"ur Theoretische Physik,\\
Freie Universit\"at Berlin, Arnimallee 14,
14195 Berlin, Germany}
\author{Verena Schulte-Frohlinde%
 \thanks{Email: frohlind@cmt.harvard.edu 
URL:
http://www.physik.fu-berlin.de/\~{}frohlind \hfil
}}
\address{Lyman Laboratory of Physics,\\
Harvard University, Cambridge, MA 02138, USA}
\maketitle
\begin{abstract}
With the help of strong-coupling theory,
we calculate
the critical exponents
of O($N$)-symmetric
$\phi^4$-theories in $4- \ep$ dimensions
up to five loops
with an accuracy comparable to that
achieved by  Borel-type
resummation methods.
\end{abstract}

%
\section{Introduction}
Recently, one of us \cite{sc,seven} has
developed a new approach to critical exponents
of field theories based on the strong-coupling limit
of variational perturbation expansions \cite{systematic,PI}.
This limit is relevant for critical phenomena if
the renormalization constants are expressed
in terms of the unrenormalized coupling constant.
The theory was first applied successfully to
 O($N$)-symmetric $\phi^4$-theories in three dimensions
yielding the three fundamental critical exponents $\nu, \eta, \omega$
with high accuracy.

The method has also been shown to work
for perturbation expansions
of these theories in
$4-\ep$ dimensions \cite{criteps}, but here only
two-loop expansions were treated, where all results
can be
written down explicitly.
In this note we want to extend these calculations to  five-loop levels.

\section{Resume of Strong-Coupling Theory}
 From model studies
of perturbation expansions of
the anharmonic oscillator
we have learned that variational perturbation expansions
possesses
good strong-coupling limits
 \cite{JK1,JK2}, with a
speed of convergence
governed
by the
 convergence
radius
of the strong-coupling expansion \cite{JK3,GZ}.
This has enabled us
to
set up
a simple
algorithm \cite{PI}
 for
deriving
uniformly
convergent
approximations to functions
of which one knows
a few initial Taylor coefficients
and an important scaling property:
the functions approach a constant value
with a given inverse power of the variable.
The renormalized coupling constant $g$
and the critical exponents
 of a $\phi^4$-theory have precisely this property
as a function of the bare coupling constant $g_B$.
In $D=4- \ep$ dimensions
 the approach is parametrized as follows  \cite{sc}
\begin{equation}
g(g_B)=g^*-\frac{{\rm const}}{g_B^{  \omega/ \ep}}+\dots~,
\label{appr}\end{equation}
where $g^*$ is the infrared-stable
fixed point, and $ \omega$ is called the critical
exponent of the approach to scaling.
This  exponent is universal, governing the approach
to scaling of every function of $ F(g)$,
\begin{equation}
f(g_B)=
 F(g(g_B)) = F(g^*) + F'(g^*) \times \frac{\rm const}{g_B}
 \equiv f^* + \frac{\rm const '}{g_B^{ \omega /\varepsilon}}.
\label{fvgB}\end{equation}
This type of scaling behavior is observed experimentally
in systems
described by $\phi^4$-theories, and strong-coupling theory
is designed to calculate $f(g^*)$ and $ \omega $.

Let $f(g_B)$ be
a function with this behavior and suppose that we
know its first $L+1$ expansion terms,
\begin{equation}
f_L(g_B)=\sum_{l=0}^L  a_l  g_B  ^l.
\label{@truncatedex}\end{equation}
More specifically than in Eq.~(\ref{appr}),
we assume that
$f(g_B)$
 approaches its constant strong-coupling limit $f^*$
in the form of an inverse power series
\begin{equation}
f_M(g_B)=  \sum_{m=0}^M b_m  (g_B ^{-2/q}) ^m ,
\label{2.4@}\end{equation}
with a finite convergence radius \cite{Interpolation}.
Then the $L$th approximation to the
value $f^*$ is obtained from the
strong-coupling
 formula \cite{sc,seven,criteps}
\begin{eqnarray}
&&f_L^* =\mathop{\rm opt}_{\hat{g}_B}\left[
\sum_{l=0}^L a_{l}^{\rm } v_l\hat{g}_B^l
       \right] ,~~~~v_l\equiv  \sum_{k=0}^{L-l}
      \left( \begin{array}{c}
              - q l/2 \\ k
             \end{array}
      \right)
     (-1)^{k} .
\label{coeffb}\end{eqnarray}
The quantities $v_l$ are simply  binomial expansions of
$(1-1)^{-ql/2}$  up to the order $L-l$.
The expression in brackets
has to be optimized in the variational parameter
$\hat g_B$.
The optimum is
the smoothest among all real extrema. If there are no real extrema,
the
turning points serve the same purpose.

The derivation  of this rule
is simple:
We replace $g_B$ in (\ref{@truncatedex}) trivially by
$\bar g_B\equiv g_B/\kappa^q$ with $ \kappa=1$.
Then we rewrite, again trivially, $ \kappa^{-q}$ as $ (K^2+ \kappa^2-K^2)^{-q/2}$
with an arbitrary parameter $K$.
Each term is now expanded in powers of $r=(\kappa^2-K^2)/K^2$
assuming $r$ to be of the order $g_B$.
Then we take the limit $ g_B\rightarrow \infty$
at a fixed     ratio
 $\hat g_B\equiv g_B/K^q$,
so that $K\rightarrow \infty$ like $g_B^{1/q}$ and
$r\rightarrow -1$,
yielding (\ref{coeffb}).
Since the final result to all orders cannot depend on the
arbitrary parameter $K$, we expect the best result to any finite order to
be optimal at an extremal value of $K$, i.e., of $\hat g_B$.

The   approach to the
limit of $r$ is
$r= -1+ \kappa^2/K^2
=-1+O(g_B^{-2/q})$.
 This implies
the leading correction
 to
$f^*_L$ to be
of the order of $g_B^{-2/q} $.
Application of the theory to a function with the
strong-coupling
behavior
(\ref{appr}) requires therefore
a parameter
 $ q = 2 \ep/ \omega$
in formula (\ref{coeffb}).
A systematic expansion in powers of $K^2$ leads to the
strong-coupling expansion (\ref{2.4@}).

For $L=2$ and $3$ we have given in Ref.~\cite{criteps}
analytic expressions
for the strong-coupling limits
(\ref{coeffb}). Setting $ \rho\equiv 1+q/2=1+ \ep/ \omega$,
the limits are for $L=2$:
\begin{eqnarray}
f_2^*= \mathop{\rm opt}_{\hat{g}_B}\left[
a_0+
a_1\rho \hat g_B +
a_2\hat g_B^2
\right] =a_0-\frac{1}{4}\frac{a_1^2}{a_2} \rho^2,
\label{f2@}\end{eqnarray}
and for $L=3$:
\begin{eqnarray}
f_3^*&=&  \mathop{\rm opt}_{\hat{g}_B}\left[
a_0+{\scriptstyle\frac{ 1}{ 2}}a_1 \rho( \rho+1)\hat{g}_B+a_2(2 \rho-1)
\hat{g}_B^2
+a_3\hat{g}_B^3 \right]        \nonumber \\
&=&a_0-\frac{1}{3}\frac{\bar a_1\bar a_2}{a_3}\left(1-\frac{2}{3}r
\right)
+\frac{2}{27}\frac{\bar a_2^3}{a_3^2}\left(1-r \right),
\label{f3@}\end{eqnarray}
where
$
r\equiv
 \sqrt{1-3{\bar a_1a_3}/{\bar a_2^2}}$ and
$ \bar a_1 \equiv {\scriptstyle\frac{ 1}{ 2}}a_1 \rho( \rho+1) $
and
$ \bar a_2 \equiv a_2(2 \rho-1) $.
The
positive
square root must be taken
  to connect $g_3^*$ smoothly to
$g_2^*$ at small  $g_B$.
If the square root is imaginary, the optimum is given by
the unique turning point, which leads once more to
the limit (\ref{f3@}), but
with $r=0$.

The exponent $ \omega$ describing the approach to scaling
can be determined from the expansion coefficients
of an arbitrary function  $h(g_B)$
behaving like
(\ref{fvgB})
as follows. Since  $h(g_B)$
goes to $h^*$ in the strong-coupling limit,
the logarithmic derivative $s(g_B)\equiv g_B h'(g_B)/h(g_B)$ must vanish at $g_B=\infty$.
If $h (g_B)$
starts out as $A_0+A_1g_B+\dots$ or as $A_1g_B+A_2g_B^2+\dots~$,
the logarithmic derivative is
\begin{eqnarray}
s (g_B)&=& A'_1 g_B+(2 A'_2- A'_1{}^2)g_B^2
\nonumber \\&&+
(A'_1{}^3-3 A'_1 A'_2 +3 A'_3)g_B^3
+\dots~,
\label{omscal}\end{eqnarray}
where $ A'_i=A_i/A_0$, or
\begin{eqnarray}
s (g_B)&=&1+\hat A_2 g_B+(2\hat A_3-\hat A_2^2)g_B^2
\nonumber \\&&+
(\hat A_2^3-3\hat A_2\hat A_3+3\hat A_4)g_B^3
+\dots~,
\label{omscal2}\end{eqnarray}
where $\hat A_i=A_i/A_1$.
The expansion coefficients on the right-hand sides
may  then be inserted into
(\ref{f2@}) or (\ref{f3@}), whose left-hand sides have to vanish
to ensure that $h(g_B)\rightarrow h^*$.

Another formula for determining $ \omega $ is based on
the fact that
if the approach $h(g_B)\rightarrow h^*$
is of the type
(\ref{fvgB}),
the function
\begin{equation}
t (g_B) \equiv g_B\frac{h''(g_B)}{h'(g_B)} =2\hat A_2 g_B
+(-4\hat A^2_2+6\hat A_3)g_B^2
+(8\hat A_2^3-18\hat A_2 \hat A_3 +12\hat A_4)g_B^3
+\dots~
\label{extraeq}\end{equation}
must have the strong-coupling limit
\begin{equation}
t(g_B)\rightarrow t^*=-\frac{ \omega} \ep-1.
\label{fome@}\end{equation}

\section{Renormalization Constants and Critical Exponents}

Let us briefly recall
the definitions
of the
$\phi^4$-theory in $D=4- \ep$ dimensions
whose five-loop expansions we want to evaluate.
The bare euclidean action is
\begin{equation}
{\cal A}\!=\!
\int\! d^Dx\left\{ \frac{1}{2}\left[\partial \phi_B(x)\right]^2\!
\!+\!   \frac{1}{2} m_B^2\phi_B^2(x)
\!+\!(4\pi)^2\frac{  \lambda_B}{4!}\left[  \phi_B^2(x)\right] ^2\!\right\}\!.
\label{@}\end{equation}
where
the field $\phi_B(x)$ is an $N$-dimensional vector,  the action being
O($N$)-symmetric.
The Ising model corresponds to $N=1$,
the superfluid phase transition by $N=2$, and the classical Heisenberg
magnet
 by $N=3$, the critical behavior of dilute polymer solutions is
described by $N=0$.

By calculating the Feynman integrals,
regularized via an expansion in $ \ep=4-D$
 and
arbitrary mass scale $\mu$,
one obtains renormalized values
of
mass, coupling constant, and field related to the bare
quantities by
renormalization constants
$Z_{\phi},Z_{m},Z_{g}$:
\begin{equation}
m_B^2
=m^2~Z_{m}Z^{-1}_{\phi},~~
   \lambda_B =  \lambda\,{Z_g}Z^{-2}_{\phi}, ~~
\phi_B =\phi\,\,Z_{\phi}^{1/2}.
\label{gefung}
\label{ren@}\end{equation}
Up to two loops,
perturbation theory yields
the following
expansions in powers of the dimensionless reduced
coupling constant
$g_B\equiv  \lambda_0/\mu^ \ep$:
\begin{eqnarray}
{ g}&=&{ g}_B -
\frac{ N+8}{3 \ep }  g_B^2 +
  \,\left[ \frac{(N+8)^2}{9 \ep^2}+\frac{3N+14}{6 \ep}\right]
{ g_B^{3}}+ \dots ~
, \label{gequ@}\\
\frac{m^2}{m_B^2}&=&1-
\frac{N+2}{3}\frac{g_B}{ \ep}
+\frac{N+2}{9}\left[\frac{N+5}{ \ep^2}+\frac{5}{4 \ep}
\right]
{g_B^2} + \dots ~,\label{gfg-0}\\
\frac{\phi^2}{\phi_B^2}&=&1+\frac{N+2}{36}\frac{g_B^2}{ \ep}
+ \dots ~
.
\label{gfg-1}\end{eqnarray}
We refrain from writing down the lengthy
five-loop expressions calculated in Ref.~\cite{KNSF},
since they can be downloaded from the internet \cite{URL}.
We now
set the scale parameter $\mu$ equal to
the physical mass $m$
and consider all quantities as functions
of
 $g_B= \lambda_B/m^ \ep$.
In order to describe second-order phase transitions,
we let $m_B^2$ go to zero like $\tau =$const$\times( T-T_c)$
as the temperature
$T$ approaches the critical temperature $T_c$, and assume
that also $m^2$ goes to zero,
and thus
$g_B$
to infinity.
The latter  assumption will be seen to be self-consistent
after Eq.~(\ref{@etamn}).
Assuming the theory to scale as suggested by experiments,
we now determine the value
of the renormalized coupling constant $g$
in the strong-coupling limit $g_B\rightarrow \infty$,
and the exponent $\omega$ of approach, assuming the behavior
(\ref{appr}). First we apply formula (\ref{coeffb}) to
the logarithmic derivative $s(g_B)$ of the function $g(g_B)$,
which is determined by Eq.~(\ref{omscal2}).
Setting $s_L^* = 0$, determines the approximation $ \omega _L$ to
$ \omega $.

The other critical exponents are found as follows
from the experimental behavior of systems described by $\phi^4$-theories.
We know that
the ratios
$m^2/m_B^2$ and $\phi^2/\phi_B^2$ have a limiting power behavior
for small $m$:
\begin{equation}
\frac{m^2}{m_B^2}\propto  g_B^{- \eta_m/ \ep}\propto m^{ \eta_m},~~~~~~
\frac{\phi^2}{\phi_B^2}\propto  g_B^{ \eta/ \ep}\propto m^{ -  \eta}.
\label{phimass}\end{equation}
The powers $\mu_m$ and $\eta$ can then be calculated
from the strong-coupling limits of the
logarithmic derivatives
\begin{eqnarray}
  \eta_{m}({ g}_B)\!=\!
- \ep\frac{d}{d\log  g_B}\log\frac{m^2}{m^2_B},~~~~
  \eta({ g_B})
\!=\!
 \ep\frac{d}{d\log  g_B}\log\frac{\phi^2}{\phi^2_B}
 .\!\!\label{etame2}
\label{etaetam@}\end{eqnarray}
Inserting (\ref{gfg-0}) and (\ref{gfg-1})
on the right-hand sides
yields the expansions
\begin{eqnarray}
 \eta _m(g_B)&=&\frac{N+2}{3 }g_B-\frac{N+2}{18}
\left(5+2\frac{N+8} \ep\right)g_B^2 + \dots ~,\label{@etam}
\\
 \eta(g_B)&=&\frac{N+2}{18}g_B^2 + \dots ~.
\label{@eta}\end{eqnarray}

When approaching the
second-order phase transitions
where the bare mass $m_0^2$
vanishes like $\tau \equiv (T-T_c)$,
the physical mass $m^2$ vanishes
 with a different power of $\tau $.
This power is obtained from the first equation
in (\ref{phimass}), which shows that
$m\propto \tau ^{1/(2- \eta_m)}$. In experiments one observes
that the coherence length of fluctuations $\xi=1/m$ increases near $T_c$
like $\tau ^{- \nu}$. Comparison with the previous equation
shows  that
 for the critical exponent
$ \nu$ is equal to $1/(2- \eta_m)$.
Similarly we see
from the second equation in
(\ref{phimass}) that
 the scaling dimension $D/2-1$ of the free field  $\phi_0$
for $T\rightarrow T_c$
is changed in the strong-coupling limit
to $D/2-1+ \eta/2$, the number $ \eta$ being the so-called anomalous dimension of the field.
This implies a change in the large-distance behavior
of the correlation functions  $\langle \phi(x)\phi(0)\rangle$
at $T_c$ from the free-field behavior
 $r^{-D+2}$ to
 $r^{-D+2- \eta}$.

The magnetic susceptibility is determined by
the integrated correlation
function $\langle \phi_B(x)\phi_B(0)\rangle$.
At zero coupling constant $g_B$, this is proportional to $1/m_B^2\propto \tau^{-1} $.
The interaction changes this to
 $m^{-2}\phi_0^2/\phi^2$. This
quantity
 has a temperature
behavior $m^{-(2- \eta)}=\tau ^{ -\nu(2- \eta)}\equiv \tau ^{- \gamma}$, which
 defines the critical exponent $ \gamma= \nu(2- \eta)$
governing the divergence of the
susceptibility line.
Using
$ \nu=1/(2- \eta_m)$ and the expansions
(\ref{@etam}), (\ref{@eta}),
we obtain for $ \gamma(g_B)$
the perturbation expansion up to second order in  $g_B$:
\begin{eqnarray}
  \gamma(g_B)&=&1+\frac{N+2}{6}{ g}_B+ \frac{ N+2 }{36}\left(N-4-2\frac{N+8}{ \ep}\right)
 g_B^2 + \dots ~
.
\label{gamma}\end{eqnarray}

\section{Explicit Two-Loop Results}
\begin{figure}[tbhp]
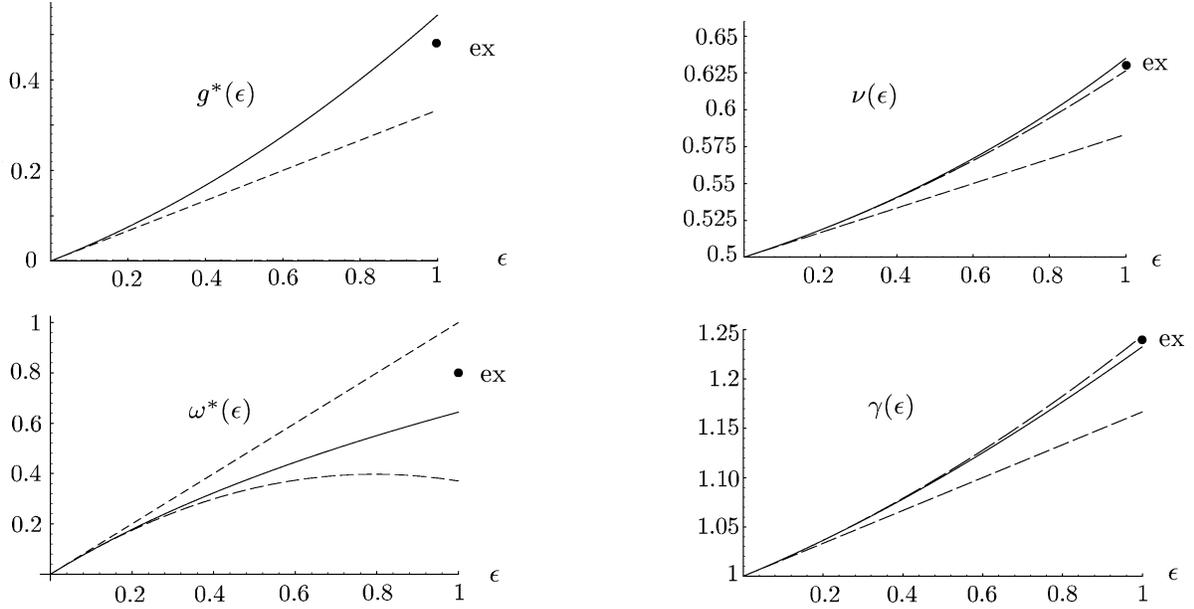

\vspace{-1cm}
\hspace{-1.5cm}
\input all.tps   ~\\
\caption[]{
Two-loop results
of strong-coupling theory
for critical exponents of
the Ising universality class $N=1$.
The first figure shows the renormalized coupling
at infinite bare coupling
as a function of $ \ep=4-D$ calculated
from the first three perturbative
expansion terms. The curve coincides with the $ \ep$-expansion
up to the order $ \ep^2$.
The dashed curve
indicates
the linear term.
The other figures show the critical exponents $ \omega$, $ \nu$, and $ \gamma$.
Dashed curves indicate linear and quadratic $ \ep$-expansions.
The dots mark the values
 $g^*\approx0.48\pm 0.003$, $ \omega\approx0.802\pm0.003$,
$ \nu=0.630\pm0.002$, and $ \gamma=1.241\pm0.004$
obtained from six-loop calculations \cite{sc}.
}
\label{omf}\end{figure}

Let us briefly recall the explicit
results obtained in Ref.~\cite{criteps}
from the above two-loop expansions.
First we
calculate
the critical exponent $ \omega$
from the
requirement that $g(g_B)$ has a constant
strong-coupling limit,
implying the vanishing of (\ref{omscal2}) for $g_B\rightarrow \infty$.
We form the logarithmic derivative
(\ref{omscal2})
of the expansion
(\ref{gequ@})
up to the order $g_B^2$, and
 obtain from Eq.~(\ref{f2@})
 the scaling condition
\begin{equation}
0=1-\frac{1}{4}\frac{\hat A_2^2}{2\hat A_3-\hat A_2^2} \rho^2
\;.
\label{ex@}\end{equation}
This fixes $ \rho = 1 + \ep/ \omega $ to
\begin{equation}
 \rho= \sqrt{8\hat A_3/\hat A_2^2-4}\;.
\label{rho@}\end{equation}
Since $ \omega$ must be greater than zero, only
the positive square root is physical.
With the explicit coefficients
$A_1,A_2,A_3$ of expansion (\ref{gequ@}), $ \rho $ becomes
\begin{equation}
 \rho=
2\sqrt{1+{3}\frac{3N+14}{(N+8)^2} \ep}  .
\label{x22@}\end{equation}
The associated critical exponent $ \omega= \ep/( \rho-1)$
is plotted in
Fig.~\ref{omf}.
It has the $ \ep$-expansion
\begin{equation}
  \omega= \ep  -3\frac{3N+14}{(N+8)^2} \ep^2+\dots~,
\label{omep@}\end{equation}
also shown in Fig. 1, which
agrees with the first two terms obtained from
renormalization group calculations \cite{KNSF}.

From Eqs.~(\ref{fome@}), (\ref{extraeq}), and (\ref{f2@}) we obtain
 for the critical exponent $  \omega$ a further equation
\begin{eqnarray}
-\frac{ \omega}{ \ep}-1= -\frac{ \rho}{ \rho-1}&=&
-\frac{1}{2}\frac{\hat A_2^2 \, \rho^2  }{3\hat A_3-2\hat A_2^2}
.
\label{@omega2}\end{eqnarray}
which is solved by
\begin{equation}
 \rho=\frac{1}{2}+ \sqrt{\frac{6\hat A_3}{\hat A_2^2}-\frac{15}{4}},
\label{@}\end{equation}
with the positive sign of the square root
ensuring
a positive
$ \omega$.
Inserting the coefficients of
(\ref{gequ@}), this becomes
\begin{equation}
 \rho=\frac{1}{2}+ \frac{3}{2} \sqrt{1+4\frac{3N+14}{(N+8)^2} \ep}.
\label{sign}\end{equation}
The associated critical exponent $ \omega= \ep/( \rho-1)$
has
the same $ \ep$-expansion (\ref{omep@})
as the previous approximation (\ref{x22@}).
When plotted in Fig.~1 the
 alternative
 approximation  (\ref{sign}) is indistinguishable from
the earlier one in (\ref{x22@}) in the plot
of Fig.~\ref{omf}.

Having determined $ \omega$, we can now calculate $g^*$.
Inserting the first two coefficients of the expansion
(\ref{gequ@}) into (\ref{f2@}) we obtain
\begin{equation}
g^*_2= a_0
-
\frac{1}{4}\frac{a_1^2}{a_2} \rho^2.
\label{@}\end{equation}
Together with (\ref{x22@}), this yields
\begin{equation}
g^*_2=\frac{3}{N+8} \ep+
9   \frac{3N+14}{(N+8)^3} \ep^2,
\label{epsex}\end{equation}
which is precisely the  $ \ep$-expansion
of $g^*$ derived
 in two-loop
renormalization group calculations.

\cm{Including the next expansion coefficient in (\ref{gequ@}),
we can use formula (\ref{f3@})
to calculate
the three loop
 approximation $g_3^*$. However, at $ \ep=1$, the square root turns out to be
imaginary, indicating that the minimum can no longer be used to
optimize the variational expression  (\ref{coeffb}).
Thus we must use the turning point
for
optimization, in which case we
obtain
the expression (\ref{f3@})
with the term
$r$ omitted.
The result
lies slightly ($\approx 8\%$) above
the curve (\ref{epsex}), i.e.,  the approximation
is not improved.  This is due to the fact that
we have not determined $ \omega $ to one more order in $\ep$.
Indeed, the $ \ep^3$-term in $g_3^*$ is
$81(3N+14)^2/8(N+8)^5$ and
disagrees in sign with the exact term
 $
{ \ep^3}[{3}\left(
 -33N^3  +110N^2+1760n +  4544\right)/8
$ $-36\zeta(3)(N+8)\left(5N+22\right)]/{(N+8)^5},$
%
which we find by calculating $ \rho$
from an expansion (\ref{gequ@}) with one more power in $g_B$.}

We now turn to the critical exponents.
Taking the expansion (\ref{@etam}) for $ \nu $ to infinite $g_B$,
we obtain from
formula (\ref{f2@}) the limiting value
\begin{equation}
 \eta_m= \frac{\ep}4 \frac{N+2}{N+8+5 \ep/2} \rho^2.
\label{@etamn}\end{equation}
This is certainly positive, so that the first equation (\ref{phimass})
ensures
that with $m_B^2$ also $m^2$ goes to zero,
a necessary condition for the
self-consistency of the theory.

The corresponding
$ \nu=1/(2- \eta_m)$
is plotted in Fig. 1.
With the approximation (\ref{x22@}) for $ \rho$,
we find for $ \nu$ the $ \ep$-expansion
\begin{equation}
 \nu=\frac{1}{2}+\frac{1}{4}\frac{N+2}{N+8} \ep+\frac{(N+2)(N+3)(N+20)}{8(N+8)^3} \ep^2+\dots~,
\label{@}\end{equation}
which is also shown in Fig.~1, and agrees with the
renormalization group result to this order in $\varepsilon$.

As a third independent
critical exponent we calculate
$ \gamma=(2- \eta)/(2- \eta_m)$ by inserting
the coefficients of the expansion (\ref{gamma})
into formula~(\ref{f2@}) and obtain
\begin{equation}
  \gamma={1}+\frac{ \ep}{8}\frac{N+2}{N+8-(N-4)\ep/{2}} \rho^2 ,
\label{@}\end{equation}
plotted in Fig.~1. This has an $ \ep$-expansion
\begin{equation}
 \gamma={1}+\frac{1}{2}\frac{N+2}{N+8} \ep
+\frac{1}{4}\frac{(N+2)(N^2+22N+52)}{(N+8)^3} \ep^2+\dots~,
\label{@}\end{equation}
also shown  in Fig.~1, and
agreeing with renormalization group results to this order.
The critical exponent
 $ \eta=
2- \gamma/ \nu$ has the $ \ep$-expansion
%
$ \eta={(N+2)} \ep^2/2{(N+8)^2}+\dots~.  $
%

The above results demonstrate
that variational
strong-coupling theory
can successfully
be applied to
 $\phi^4$-theories
in $D=4- \ep$ dimensions
and yields resummed expressions
for the $ \ep$-dependence
of all  critical exponents.
Their $ \ep$-expansions agree with
those obtained from renormalization group
calculations.

In order to achieve better accuracies
we shall now apply the method
to five-loop expansions.

\section{Extension to Five Loops}
The above calculations are  again
extended
to five-loops using the power series for
the critical exponents of Ref.~\cite{KNSF,URL}.
In a first step,
we determine the parameter $\rho=1+\ep/\omega$
for which
 the logarithmic derivative of $g(g_B)$ approaches zero for
$g_B\rightarrow \infty$ [see Eq.~(\ref{omscal2})].
We therefore insert the coefficients of
the power series of $s(g_B) \equiv
g_B g'(g_B)/g(g_B)$
  into Eq.~(\ref{coeffb}) and determine
$\rho = 1 + \ep/ \omega $ for $L=2,3,4,5$,
to make $s^*_L=0$.
The resulting  $\ep$-expansions for the approach-to-scaling parameter $\omega$ reproduce the
well-known $\ep$-expansions
in \cite{KNSF} up to the corresponding order.
In Figure~\ref{figomalt}, the approximations $\omega_L$
are plotted against the number of loops $L$ for $\ep=1$.

Apparently, the five loop results are still some distance away from
a constant $L \rightarrow \infty$-limit.
The slow  approach to the  limit calls
for a suitable extrapolation method.
The  general convergence behavior in the
limit
$L\rightarrow \infty$
 was determined in \cite{sc} to be of the general form
\begin{equation}
f^*(L) \approx f^*+{\rm const}\times e^{-c\,L^{1-\omega}}.
\label{funcform}
\end{equation}
We therefore plot the approximations $s_L$
for a given $ \omega $ near the expected critical exponent
against $L$.
To exploit this knowledge
and fit the points by the theoretical curve (\ref{funcform})
 to determine the limit $s^*$.
Then $ \omega $ is varied, and the plots are repeated until $s^*$
is zero. The resulting $\omega$
is the desired critical exponent, and the associated plots
are shown in Fig.~\ref{@figapprom}.
Since the optimal variational parameter $\hat g_B$
comes from minima and turning points
for even and odd approximants in alternate order,
the points are best fitted by two different curves.



\cm{A faster convergent procedure was developed and applied successfully in
     Ref.~\cite{seven}.
    }
In order to determine the common constant $c$
one plots even and odd approximations $s_L$
     directly against
the variable $x_L = e ^{-cL^{1- \omega }}$.
The constant $c$ is then used to fit  straigt lines
through even and odd approximations which cross at zero $x_L$.
This procedure is shown in Fig.~\ref{fitom},
and yields the curve shown in Fig.~\ref{figom}.
The resulting $ \omega $-values are listed in Table~\ref{@criticalpap}.
They
will now be used
to derive the strong-coupling limits
for the exponents $\nu$, $\gamma$ and $\eta$.

\begin{figure}[tb]
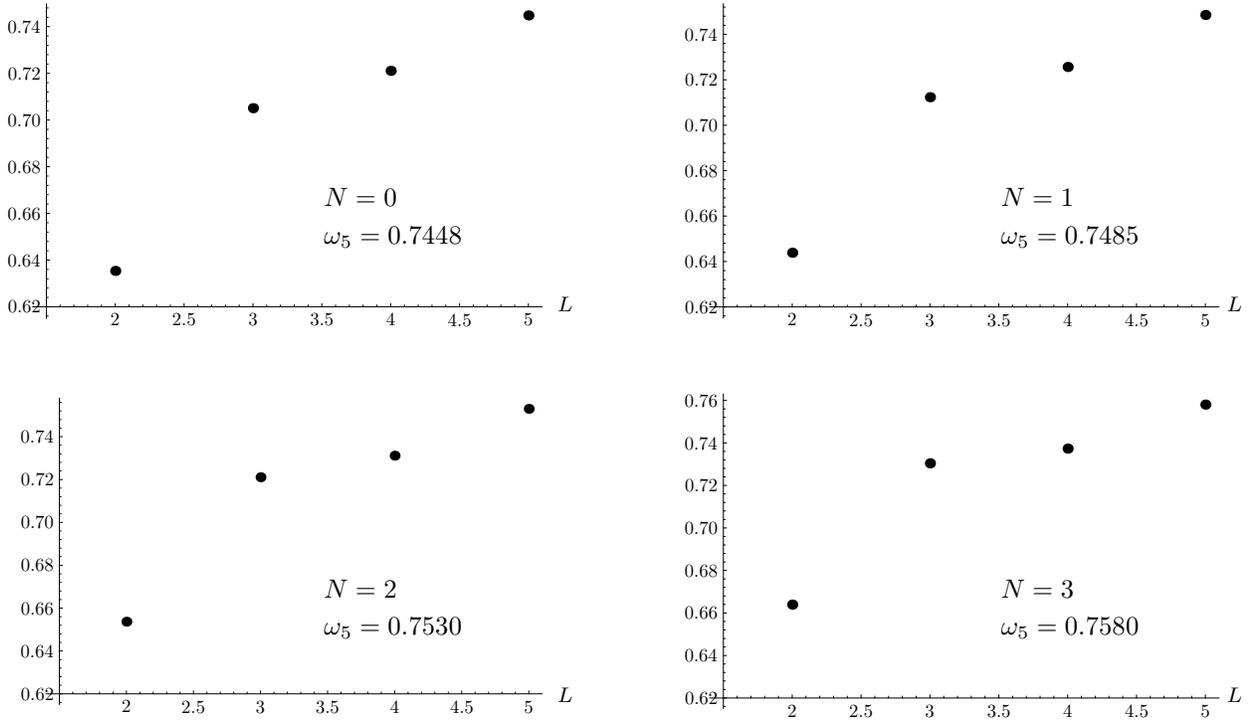

\input altomn.tps
\label{figomalt}
\caption[]{Critical exponent of approach to scaling $\omega$ calculated from
$s^*_L=0 $, plotted against the order of approximation $L$.}
\label{figomalt}\end{figure}

\noindent
\begin{figure}[bp]
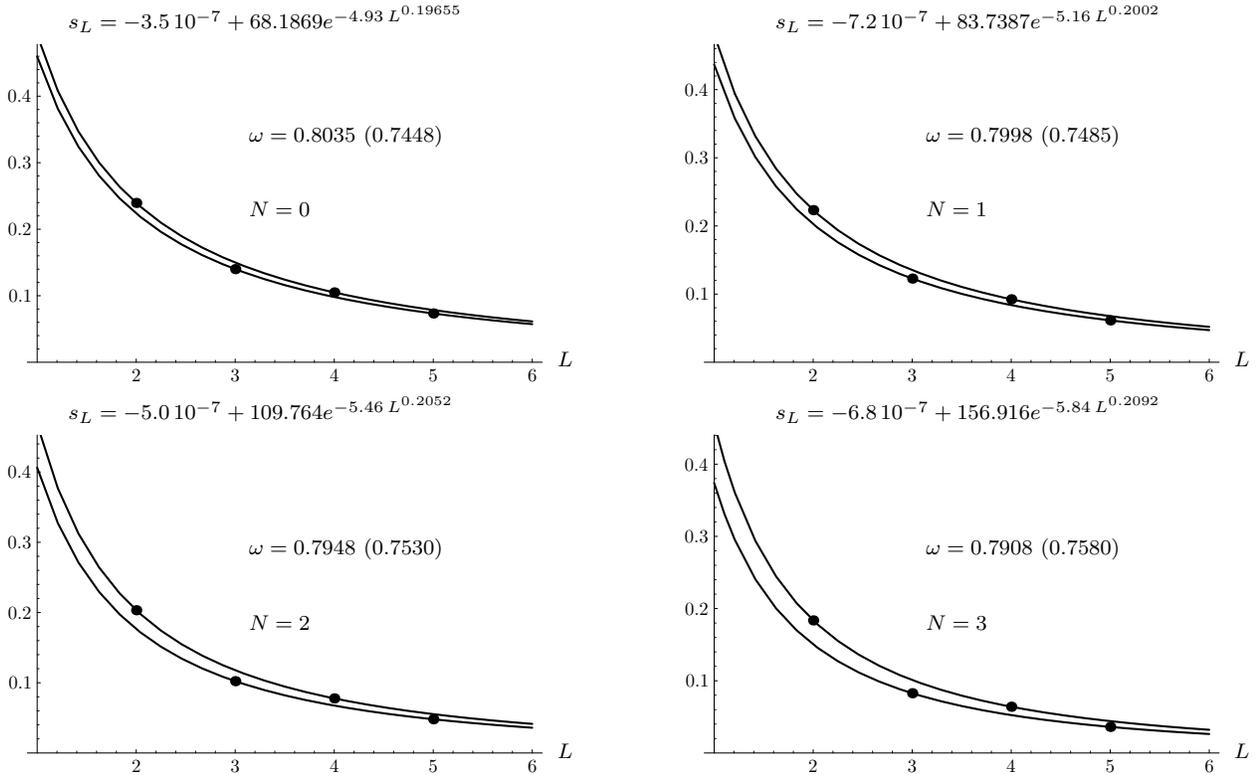

\input fitomn.tps
\caption[]{\label{fitom}Extrapolation of the solutions of the
equation $ s^*_L =0$  to $L\rightarrow \infty$ with the help of the theoretically
expected large-$L$ behavior (\protect\ref{funcform}).
The $ \omega $ where $s^*_L$ goes to zero for
$L\rightarrow \infty$ determines the critical exponent $\omega=2/q$.
The best extrapolating function is written on top of the figure.}
\label{@figapprom}\end{figure}
\begin{figure}[htp]
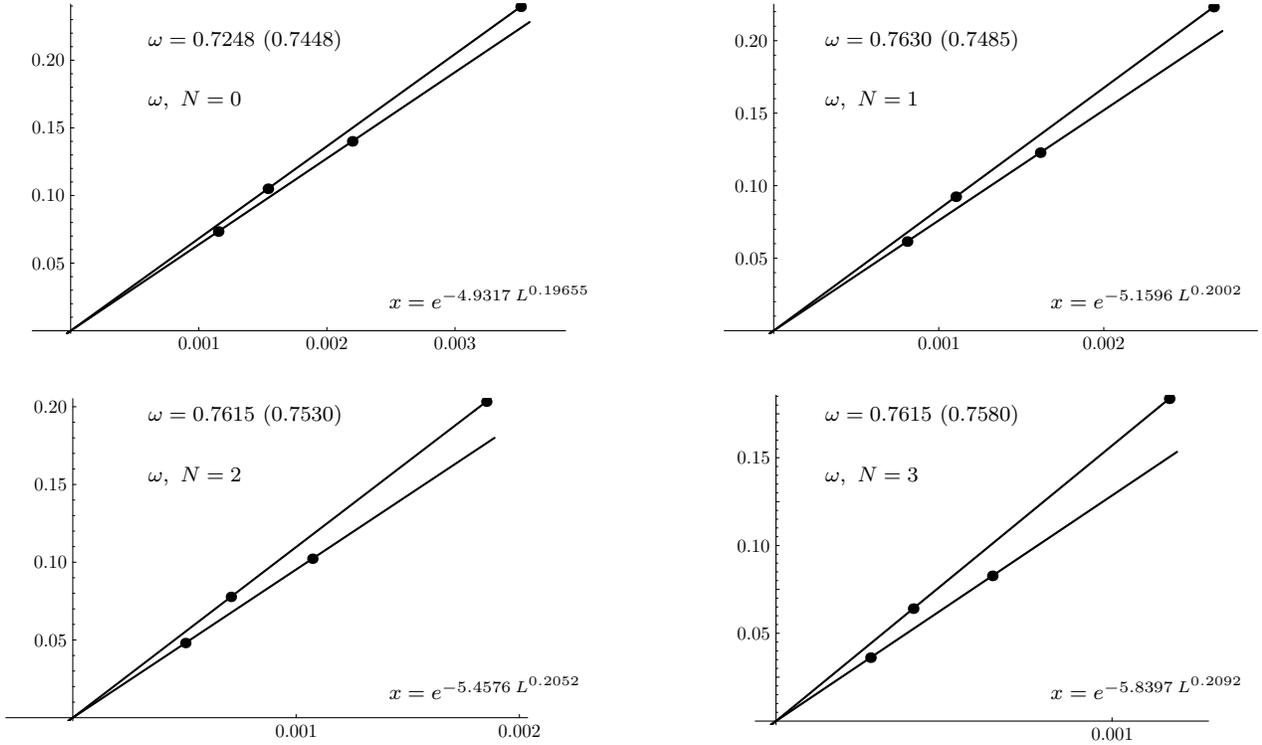

\input figomn.tps
\caption[]{\label{figom}Same plot as in Fig.~\protect\ref{@figapprom},
but  against the
variable $x_L=e^{-cL^{1- \omega }}$.
The parameter $c$ is fixed by
requiring the straight lines to cross on the vertical axis.
When  this intercept lies at the origin,
we have found
 the critical exponent $ \omega $, written on the top of each plot.
For comparison, we also show a direct plot against $L$ in
Fig.~\protect\ref{@figapprom}.}
\end{figure}
%

\subsection{Exponents $ \nu$ }
\begin{figure}[btp]
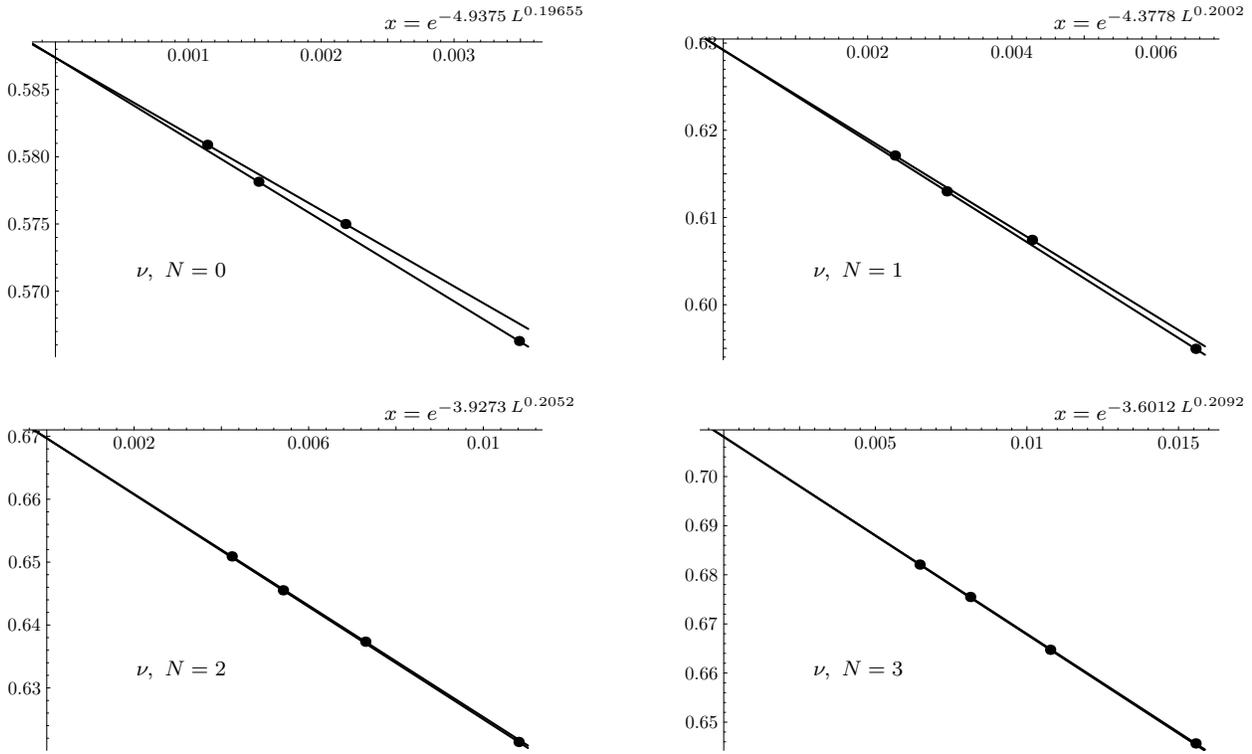

\input fignun.tps
\caption[]{\label{nufig}Critical exponent $\nu_L$(I) obtained from variational perturbation theory
plotted as a function of $x_L$.
Requiring the lines to cross at $x_L=0$ determines
the parameter $c$ in $x_L$.
See in the text.}
\end{figure}

\begin{figure}[tp]
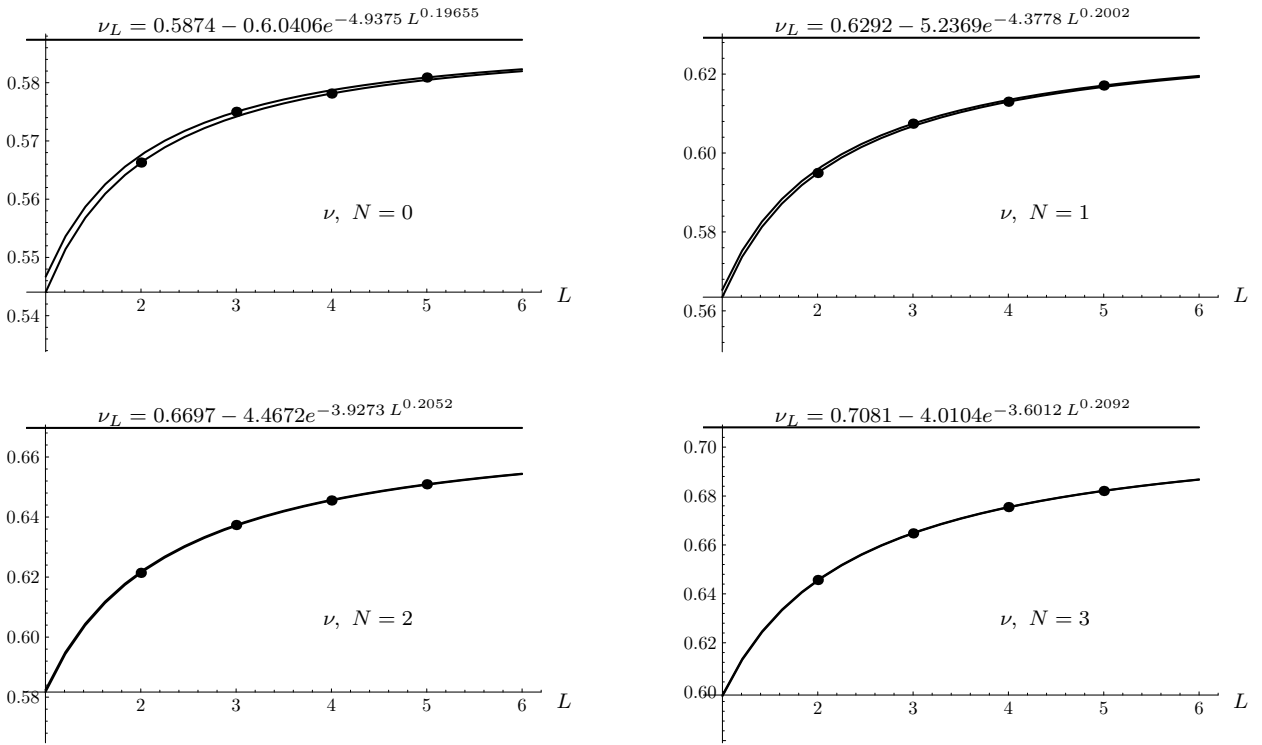

\input fitnun.tps
\caption[]{\label{nufit}Same plot as in Fig.~\ref{nufig},
but against $L$.
The fit-function is written on top of the figure.}
\end{figure}

For the calculation of the critical exponent $\nu$,
we proceed in two different ways.
This will
give us an idea of the
systematic error of the method.
 First we find the five-loop expansions for $ \nu (g_B) $
using the relation
$\nu(g_B)=1/[2-\eta_m(g_B)]$.
From this we calculate their strong-coupling
approximation $ \nu _L$
  for $L=2,3,4,5$.
After extrapolating  these to infinite $L$,
we obtain the numbers
listed
for different universality classes O($N$)
 in Table~\ref{@criticalpap} under the heading (I).
The corresponding extrapolation fits are
 plotted in Figure~\ref{nufig} and \ref{nufit}.
The resulting values for the critical
exponent $\nu(\infty)$ are indicated by
horizontal lines in RFig.~\ref{nufit}.

The second way proceeds by
calculating the  strong-coupling values
of $\eta_m(g_B)$ for $L=2,3,4,5$.
After extrapolating  these to infinite $L$,
the critical exponent $ \nu $ is
found from
$\nu=1/(2-\eta^*_m)$.
The results
are listed
 in Table~\ref{@criticalpap} under the heading (II).
The table shows in parentheses
the $L=5$ approximation for each quantity, from
which we see the extrapolation distance
of this value from the infinite-$L$ limit.

By repeating all calculations
for a slightly different $ \omega $-value, we deduce the
dependence of our results on the critical exponent $\omega$
used in the resummation process:
\begin{equation}
 \Delta  \nu
 =\left\{
\begin{array}{r}
-0.0900\times ( \omega -0.8035)\\
-0.1375\times ( \omega -0.7998)\\
-0.1853\times ( \omega -0.7948)\\
-0.2271\times ( \omega -0.7908)
\end{array}\right\}
~~~~~
{{\rm for}}       ~~~~~
\left\{ \begin{array}{l}
N=0\\
N=1\\
N=2\\
N=3
\end{array}\right\} .%
\label{Critcalex}\end{equation}

\subsection{Exponents $  \eta $ and  $ \gamma  $}
\begin{figure}[tp]
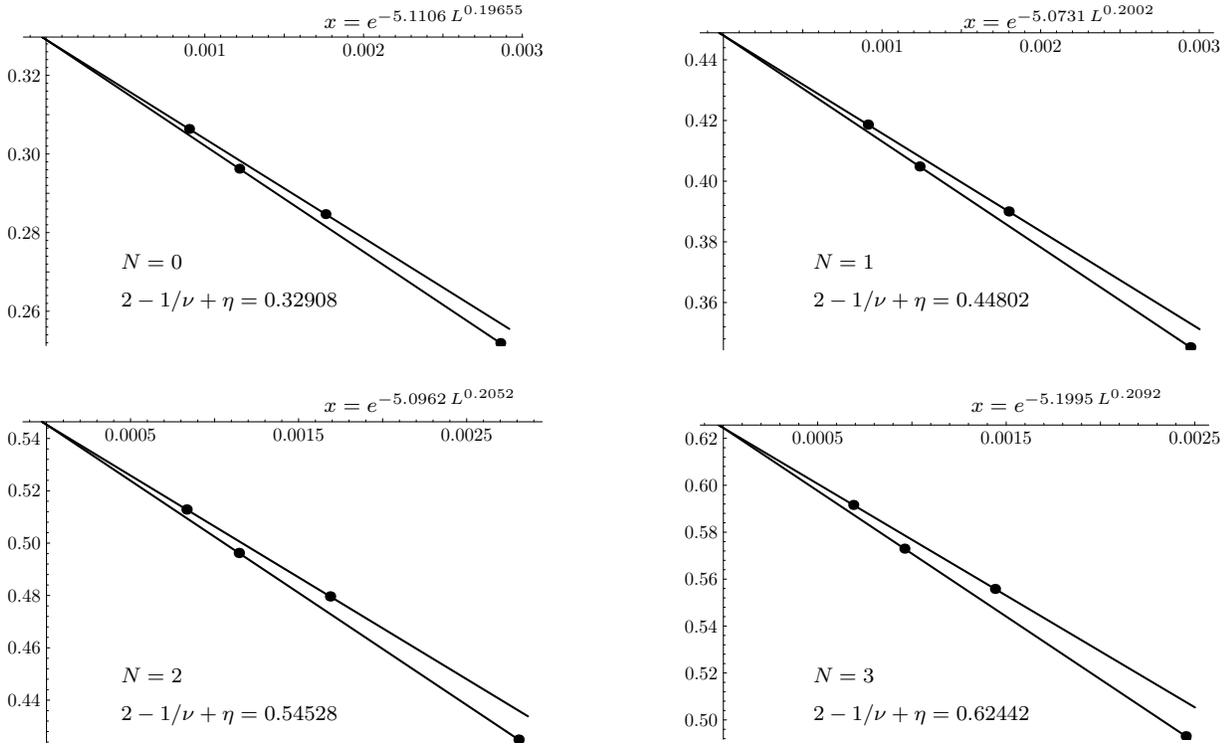

\input figetan.tps
\caption[]{\label{etafig}Determination of the critical exponent $\eta$
from the strong-coupling limit of $\eta_m+\eta$
plotted as a function of $x_L$.
Requiring the lines to cross at $x_L=0$ determines
the parameter $c$ in $x_L$.
See in the text.}
\end{figure}

\begin{figure}[tp]
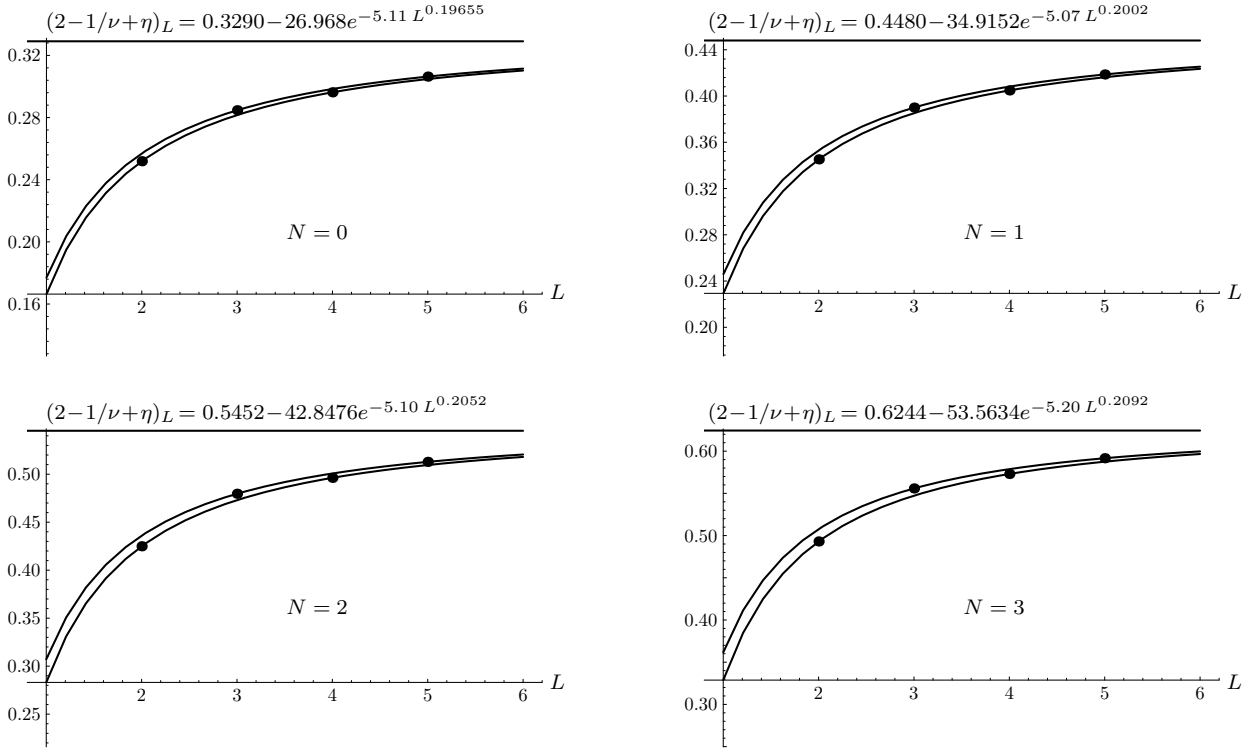

\input fitetan.tps
\caption[]{The same as above, plotted against
the order of approximation $L$ and
the fit-function written on top of the figure.
}
\label{etafit}\end{figure}

The calculation of the critical exponent $\eta$ is difficult
in all resummation schemes
since the power series of $\eta(g_B)$ starts out with
$g_B^2$,
so that there is one approximation less than for $ \nu $.
The  three points approximation $\eta_3, \eta_4, \eta_5$
 we obtain from the
five-loop expansions are not sufficient to carry out the above
extrapolation procedure.
The exponent is therefore calculated from the strong-coupling limit
of the power series for
$\bar  \eta (g_B)\equiv
\eta_m(g_B)+\eta(g_B)$
which supplies us with the combination of critical
exponents $2-1/\nu+\eta$.
After finding $\bar \eta^*$   we
subtract from this $2-1/\nu$ and obtain the desired $ \eta $.
If we use
$\nu$ (I)
of Table~\ref{@criticalpap}
in this subtraction,
we obtain $ \eta $-values listed as
$\eta$ (I)
in Table~\ref{@criticalpap}.
From
$\nu$ (II) we get $\eta$ (II).
The fits leading to the strong-coupling limits of $\bar  \eta (g_B)$
are shown in Figures~ \ref{etafig} and \ref{etafit}.
As before, the limiting values
for $L\rightarrow \infty$ are indicated by
horizontal lines.
The fitted extrapolation function is displaed on top of each figure.

An independent strong-coupling calculation for
the critical exponent $ \eta $
may be obtained
by resumming the
 series expansion
for the critical exponent of the susceptibility $\gamma=\nu(2-\eta)$.
The extrapolation plots for this exponent are shown in
Figs.~\ref{gamfit} and \ref{gamfig}.
The resulting value for $\gamma$ is also contained in Table~\ref{@criticalpap}.
As in all entries, we have listed
the fifth-order approximations in parentheses
to illustrate the extrapolation distance
to infinite order $L$.

The dependence on the value of $\omega$ is of the same order
of magnitude as for $\nu$:
\begin{equation}
 \Delta  \gamma
 =\left\{
\begin{array}{r}
-0.1500\times ( \omega -0.8035)\\
-0.2237\times ( \omega -0.7998)\\
-0.3147\times ( \omega -0.7948)\\
-0.4014\times ( \omega -0.7908)
\end{array}\right\}
~~~~~
{{\rm for}}       ~~~~~
\left\{ \begin{array}{l}
N=0\\
N=1\\
N=2\\
N=3
\end{array}\right\} .%
\label{@}\end{equation}

\subsection{Comparison with Previous Results and Experiments}

In Table \ref{@criticalpap} we have added to our results
also those
obtained by other methods. Since an extensive Table has been
published before
(Table~IV in Ref.~\cite{seven}),
we confine ourselves here only to
results of the resummation of the $\ep$-expansion
by Guida and Zinn-Justin in \cite{GZ}, and those
from three-dimensional variational perturbation
theory to sixth order  for $ \omega $ and to order 7 for $ \nu $ and $ \eta$
in Ref.~\cite{seven}.
The difference between $\nu$ (I) and $\nu$ (II), and
$\eta$ (I) and $\eta$ (II) is considerably smaller than the typical
errors in the other references.

The results of our strong-coupling theory
agree very well with those obtained from  Borel-type resummation
although we do not make use of the known large-order behavior.
For a good test of the reliability of our results we compare
  our results with
experiments.
The most precise experimental values are available from specific heat
measurements
performed on superfluid helium near the $ \lambda $-point
at zero gravity in the space shuttle
in 1992, which are reported in Ref.~\cite{rLipa}.
There one finds for the essential exponent $ \alpha =2-3 \nu $
the value
\begin{equation}
 \alpha =-0.01285\pm0.00038,
\label{@}\end{equation}
corresponding to
\begin{equation}
 \nu  =0.67095\pm0.00013.
\label{@}\end{equation}
Our resummation results in Table~\ref{@criticalpap}
imply a value
\begin{equation}
 \nu _{\rm ours}=0.6697\pm0.0013,
\label{@}\end{equation}
corresponding to
\begin{equation}
 \alpha  _{\rm ours}=-0.0091\pm0.0039.
\label{@}\end{equation}
This agrees satisfactorily with
the experimental result.
In Fig.~\ref{fig11} we have compared our result with
other experiments and various theoretical determinations.
\subsection{Conclusion}
Application of strong-coupling theory
to fix-loop perturbation expansions of O($N$)-symmetric
$\phi^4$-theories in 4-$\ep$ dimensions yield satisfactory
values for all critical exponents.

%
%

%


\begin{figure}[tp]
\input figgamn.tps
\caption[]{\label{gamfig}Critical exponent $\gamma$
obtained from variational perturbation theory
plotted as a function of $x_L$.
Requiring the lines to cross at $x_L=0$ determines
the parameter $c$ in $x_L$.
See in the text.}
\end{figure}
\begin{figure}[tp]
\input fitgamn.tps
\caption[]{\protect\label{gamfit} Same plot as in Fig.~(\protect\ref{gamfig}),
but against $L$.
The fit-function is written on top of the figure.}
\end{figure}%

\begin{figure}[tp]
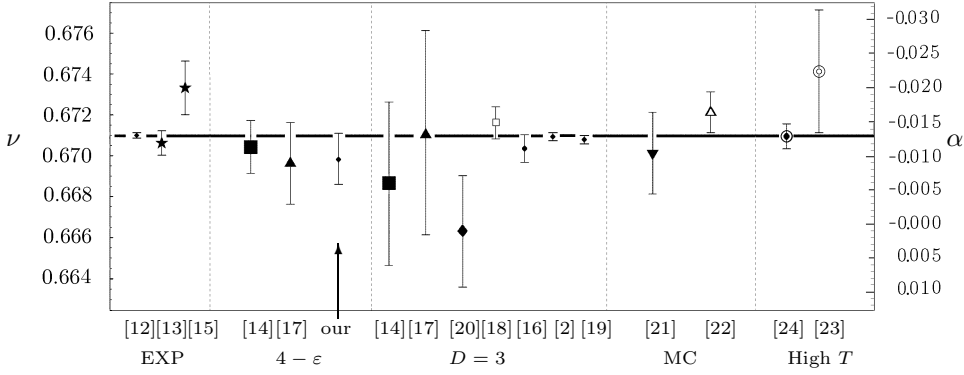

~~~~~~~~~~~~~\input nu2my.tps
\caption[Critical exponent $  \nu  $
in comparison with experimental data and results from
other resummation schemes]{Critical exponent $  \nu  $
in comparison with experimental data and results from
other resummation schemes.\label{fig11}}
\end{figure}%

\begin{table}
\begin{center}
\begin{tabular}{|c|cc|c||l|l|}
&VPT,~ $D=4-\ep$&&\hspace{-3cm}Borel-Res.~(GZ) &\hspace{-2.3cm} VPT, $D=3$ &\hspace{-2.3cm} MN, $D=3$ \\
\hline
& $\omega^{-\infty}(\omega_5)$ &             & &                   & \\
\hline
$N=0$ &$\!\!\!\!\!\!\!\!$ ~0.80345(0.7448) & &\hspace{-3.3cm} 0.828 $\pm$ 0.023 & \hspace{-2.1cm} 0.810 \hspace{-3.1cm}  &   \\
$N=1$ &$\!\!\!\!\!\!\!\!$ 0.7998(0.7485)  & & \hspace{-3.3cm} 0.814 $\pm$ 0.018 & \hspace{-2.1cm} 0.805  \hspace{-3.1cm} &  \\
$N=2$ &$\!\!\!\!\!\!\!\!$ 0.7948(0.7530)  & & \hspace{-3.3cm} 0.802 $\pm$ 0.018 & \hspace{-2.1cm} 0.800  \hspace{-3.1cm} &  \\
$N=3$ &$\!\!\!\!\!\!\!\!$ 0.7908(0.7580)  & & \hspace{-3.3cm} 0.794 $\pm$ 0.018 & \hspace{-2.1cm} 0.797  \hspace{-3.1cm} &
\\
\hline
\hline
      &\hspace{-3cm}  $ \nu^{-\infty}(\nu_5)~$ (I)&\hspace{-3cm}  $\nu^{-\infty}(\nu_5)~$ (II)&& &   \\
\hline
$N=0$ &\hspace{-3cm} 0.5874(0.5809) &\hspace{-3cm}  0.5878(0.5832) &\hspace{-3cm} 0.5875 $\pm$ 0.0018 &\hspace{-2cm}0.5883 \hspace{-3.1cm} &$ \!\!$$ \!\!$ \hspace{-2.5cm} 0.5872 $\pm$ 0.0004\\
$N=1$ &\hspace{-3cm} 0.6292(0.6171) &\hspace{-3cm}  0.6294(0.6222) &\hspace{-3cm} 0.6293 $\pm$ 0.0026 &\hspace{-2cm}0.6305  \hspace{-3.1cm}&  \hspace{-2.5cm}0.6301 $\pm$ 0.0005\\
$N=2$ &\hspace{-3cm} 0.6697(0.6509) &\hspace{-3cm}  0.6692(0.6597) &\hspace{-3cm} 0.6685 $\pm$ 0.0040 &\hspace{-2cm}0.6710  \hspace{-3.1cm}&$ \!\!$  $ \!\!$\hspace{-2.5cm} 0.6715 $\pm$ 0.0007\\
$N=3$ &\hspace{-3cm} 0.7081(0.6821) &\hspace{-3cm}  0.7063(0.6951) &\hspace{-3cm} 0.7050 $\pm$ 0.0055 &\hspace{-2cm}0.7075  \hspace{-3.1cm}&$ \!\!$$ \!\!$ \hspace{-2.5cm} 0.7096 $\pm$ 0.0008\\
\hline
\hline
       &\hspace{-3cm}  $\eta^{-\infty}(\eta_5)~$(I) &\hspace{-3cm}  $\eta^{-\infty}(\eta_5)~$(II)& & &\\
\hline
$N=0$ & \hspace{-3cm}0.0316(0.0234) &\hspace{-3cm} 0.0305(0.0234) &\hspace{-3cm} 0.0300 $\pm$ 0.0060  &\hspace{-2cm}0.03215\hspace{-3.1cm} &$ \!\!$$ \!\!$ \hspace{-2.5cm} 0.0297 $\pm$ 0.0009\\
$N=1$ & \hspace{-3cm}0.0373(0.0308) &\hspace{-3cm} 0.0367(0.0308) &\hspace{-3cm} 0.0360 $\pm$ 0.0060  &\hspace{-2cm}0.03370 \hspace{-3.1cm}&  \hspace{-2.5cm}0.0355 $\pm$ 0.0009 \\
$N=2$ & \hspace{-3cm}0.0396(0.0365) &\hspace{-3cm} 0.0396(0.0365) &\hspace{-3cm} 0.0385 $\pm$ 0.0065  &\hspace{-2cm}0.03480 \hspace{-3.1cm}&  \hspace{-2.5cm}0.0377 $\pm$ 0.0006\\
$N=3$ & \hspace{-3cm}0.0367(0.0409) &\hspace{-3cm} 0.0402(0.0409) &\hspace{-3cm} 0.0380 $\pm$ 0.0060  &\hspace{-2cm}0.03447 \hspace{-3.1cm}&$ \!\!$$ \!\!$ \hspace{-2.5cm} 0.0374 $\pm$ 0.0004\\
\hline
\hline
        &$\gamma^{-\infty}(\gamma_5)$& &   &         &\\
\hline
$N=0$   & 1.1576(1.1503)  & \hspace{-1cm}&\hspace{-3cm} 1.1575 $\pm$ 0.0050  &\hspace{-2cm}1.616\hspace{-3.1cm} &$ \!\!$\hspace{-2.5cm} 1.1569 $\pm$ 0.0004\\
$N=1$   & 1.2349(1.2194)  & \hspace{-1cm}&\hspace{-3cm} 1.2360 $\pm$ 0.0040  &\hspace{-2cm}1.241 \hspace{-3.1cm}&  \hspace{-2.5cm}1.2378 $\pm$ 0.0006\\
$N=2$   & 1.31045(1.2846) & \hspace{-1cm}&\hspace{-3cm} 1.3120 $\pm$ 0.0085  &\hspace{-2cm}1.318 \hspace{-3.1cm}&  \hspace{-2.5cm}1.3178 $\pm$ 0.0010\\
$N=3$   & 1.3830(1.3452)  & \hspace{-1cm}&\hspace{-3cm} 1.3830 $\pm$ 0.0135  &\hspace{-2cm}1.390 \hspace{-3.1cm}&$ \!\!$\hspace{-2.5cm} 1.3926 $\pm$ 0.0010\\
\end{tabular}
\end{center}
\caption[Critical exponents of five-loop strong-coupling theory
and comparison with other results]
{Critical exponents  of five-loop strong-coupling theory and
comparison with the results
from Borel-type resummation of Refs.~\cite{GZ} (GZ) and \cite{MN} (MN),
and from variational perturbation theory of Ref.~\cite{seven}.
The parentheses behind each number show the five-loop approximation
to see the extrapolation  distance.
The two values for $ \nu $ come once from a resummation  of the series for $ \nu$
itself (I), once from the series for $ \nu ^{-1}$ (II).
The two values for $ \eta $ come from subtracting once the value
$\nu(I)$ and once the value $\nu(II)$.
}
\label{@criticalpap}\end{table}

\end{document}